\documentstyle[12pt]{article}
\begin{document}
\thispagestyle{empty}
\begin{center}
\LARGE \tt \bf{Riemannian conical 2-D geometry of Heisenberg ferromagnets}
\end{center}
\vspace{1cm}
\begin{center}
{\large L.C.Garcia de Andrade }{\footnote{Departamento de Fisica Teorica-IF-UERJ-e-mail:garcia@dft.uerj.br}}\\
\end{center}
\vspace{0.5cm}
\begin{abstract}
An exact 2-dimensional conical Riemannian defect solution of 3-dimensional Euclidean Einstein equations of stresses and defects representing a shear-free Heisenberg ferromagnet is given.The system is equivalent to the Einstein equations in vacuum.Geodesics of magnetic monopoles around the ferromagnet are also investigated.
\end{abstract}
\vspace{0.5cm}
\begin{center}
\Large{PACS number(s):0450}
\end{center}
\newpage
\pagestyle{myheadings}
\markright{\underline{Riemannian 2-D conical geometry of Heisenberg ferromagnets}}
\paragraph*{}
\section{Introduction}
Earlier H.Kleinert \cite{1} showed that it is possible to obtain a 3-D Euclidean Einstein equations to investigate the geometric theory of stresses and defects in condensed matter physics \cite{2}.This equation would be written as  
\begin{equation}
G_{ij}={\sigma}_{ij}
\label{1}
\end{equation}
where ${\sigma}_{ij}$ (i,j=1,2,3) would be the stress tensor and $G_{ij}=R_{ij}-\frac{1}{2}{\delta}_{ij}R$ is the Euclidean Einstein tensor in 3-D,where $R_{ij}$ is the Ricci tensor and $R$ is the Ricci scalar.In this letter we show that it is possible to obtain an exact Riemannian conical defect solution of the Einstein vacuum equation $R_{ik}=0$ for a shear-free Heisenberg ferromagnet.As pointed out by Pismen \cite{2}:"In a 3-D isotropic (amorphous) Heisenberg ferromagnet the order parameter space,allowing for an imperfect orderering,is the interior,of the unit ball,and the minimal manifold is the 2-sphere $S^{2}$".Making use ofpolar and azimuthal angles ${\alpha}$ and ${\beta}$ to parametrize the orientation sphere ,the 3-D metric tensor can be given by \cite{2}
\begin{equation}
g_{ij}=diag(k^{2}({\rho}),{\rho}^{2},{\rho}^{2}sin^{2}{\alpha})
\label{2}
\end{equation}
where the first element concerns the energetic costs of the spatial variations of the modulus of the order parameter.The value $k=1$ yields the Euclidean metric tensor in spherical coordinates.Pismen also pointed out that the common crystalline ferromagnets are anisotropic and have prefered 
directions of easy magnetization.When there are two prefered directions,the defects are domain walls.Since the Riemannian geometry of domain walls have been investigated in detail lately \cite{3,4} it seems worth while to investigate the Riemannian geometry of these ferromagnets.By the way S.Amari \cite{5} for the anisotropy reasons mentioned above, have proposed many years ago a model for ferromagnets and Bloch walls based on Finsler Geometry \cite{6}.More recently I have proposed \cite{7} a Teleparallel model for Bloch walls where Cartan torsion appears as a Dirac delta distribution.In Amari Finslerian model two types of torsion appears, the first is related to dislocations inside the ferromagnet and the second related to the spin.Of course since the ferromagnets here possess polarized spins a more natural theory to address these issues would be a non-Riemannian theory with torsion 
and not a Riemannian manifold.But we must remenber that our solution is a vacuum solution and since we are not considering spin waves or torsion propagation in vacuum the most natural geometry to investigate this problem would be the Riemann geometry.   
\section{Riemannian Ferromagnets}
In this section we shall be concerned with the Cartan calculus formalism of differential forms to be used in this paper and also the solution of the Euclidean 3-D for the proposed source.Thus let us consider the first and second Cartan's structure equation respectively as
\begin{equation}
T^{i}=de^{i}+{\omega}^{i}_{j}{\wedge}e^{j} 
\label{3}
\end{equation}
where $e^{i}$ is the basis one-form and ${\omega}^{i}_{j}$is the connection one-form,and
\begin{equation} 
R^{i}_{j}=R^{i}_{jkm}e^{k}{\wedge}e^{m}=d{\omega}^{i}_{j}+{\omega}^{i}_{k}{\wedge}{\omega}^{k}_{j}
\label{4}
\end{equation}
Here ${\wedge}$ is the exterior product of forms symbol and $R^{i}_{jkl}$ are the components of the Riemann curvature tensor.Rewriting the metric (\ref{2}) in the differential forms language one obtains
\begin{equation}
ds^{2}={\eta}_{ij}e^{i}e^{j}
\label{5}
\end{equation}
where ${\delta}_{ij}=diag(+1,+1,+1)$ is the tetradic Kronecker delta.The basis one-form are given by 
\begin{equation}
e^{1}=k({\rho})d{\rho}
\label{6}
\end{equation}
and 
\begin{equation}
e^{2}={\rho}d{\alpha}
\label{7}
\end{equation}
and the last one\begin{equation}
e^{3}={\rho}sin{\alpha}d{\beta}
\label{8}
\end{equation}
By substitution of expressions above into expression (\ref{4}) one obtains
\begin{equation}
T^{1}={\rho}{\omega}^{1}_{2}{\wedge}d{\alpha}+{\rho}{\omega}^{1}_{3}{\wedge}sin{\alpha}d{\beta} 
\label{9}  
\end{equation}
and
\begin{equation}
T^{2}=[k{\omega}^{3}_{1}-sin{\alpha}d{\beta}]{\wedge}d{\rho}  
\label{10}  
\end{equation}
and
\begin{equation}
T^{3}=[k{\omega}^{2}_{1}-d{\alpha}]{\wedge}d{\rho}
\label{11}
\end{equation}
Since we are dealing here with the torsion-free Riemannian geometry a simple way to compute the Riemann curvature tensor components is to make the torsion components above to vanish.Thus the vanishing of both $T^{2}$ and $T^{3}$ yields the following expression for the connection one-forms
\begin{equation}
{\omega}^{3}_{1}=\frac{sin{\alpha}}{k}d{\beta}
\label{12}
\end{equation}
and 
\begin{equation}
{\omega}^{2}_{1}=\frac{d{\alpha}}{k}
\label{13}
\end{equation}
Substitution of expressions (\ref{12}) and(\ref{13}) into $T^{1}$ makes this last component of torsion two-form to vanish identically.By substitution of these same expressions into the relation (\ref{4}) allow us to write the components of the curvature two-forms $R^{i}_{j}$ as
\begin{equation}
R^{3}_{1}=d{\omega}^{3}_{1}
\label{14}
\end{equation}
and the remaing ones as
\begin{equation}
R^{3}_{2}={\omega}^{3}_{1}{\wedge}{\omega}^{1}_{2}
\label{15}
\end{equation}
and finally
\begin{equation}
R^{2}_{1}=d{\omega}^{2}_{1}    
\label{16}
\end{equation}
from these expressions we obtain the following components of the Riemann tensor
\begin{equation}
R^{2}_{113}=\frac{sin{\alpha}k'}{k^{2}}
\label{17}
\end{equation}
and
\begin{equation}
R^{2}_{113}=\frac{-cos{\alpha}}{k}
\label{18}
\end{equation}
and 
\begin{equation}
R^{3}_{112}=\frac{k'}{k^{2}}
\label{19}
\end{equation}
and the last non-vanishing component reads 
\begin{equation}
R^{3}_{213}=\frac{sin{\alpha}}{k^{2}}
\label{20}
\end{equation}
One notes from these expressions that the Riemann components vanish (Euclidean case)in the when the order parameter $k$ approaches infinity.Physically this means that the Heisenberg ferromagnet acts as a geometrical defect.
Substitution of the last expressions in the Euclidean vacuum Einstein equations reduces to the shear components of the stress tensor 
\begin{equation}
R_{12}=\frac{sin{\alpha}}{k^{2}}={\sigma}_{12}=0
\label{21}
\end{equation}
and
\begin{equation} 
R_{23}=-\frac{cos{\alpha}}{k}+(1+sin{\alpha})\frac{k'}{k^{2}}={\sigma}_{23}
\label{22}
\end{equation}
A simple 2-D solution of Einstein Euclidean equations for the ferromagnets can be obtained by noticing that ${\sigma}_{13}=0$ leads $sin{\alpha}=0$,
${\alpha}=0$ and ${\sigma}_{23}=0$ leads to the following expression 
\begin{equation} 
k'-k=0
\label{23}
\end{equation}
The simple differential equation yields the following solution
\begin{equation}
k({\rho})=e^{{\rho}+c}
\label{24}
\end{equation}
where $c$ is an integration constant.The Heisenberg ferromagnet 2-D,metric is 
\begin{equation}
ds^{2}=e^{{\rho}+c}d{\rho}^{2}+{\rho}^{2}d{\beta}^{2}
\label{25}
\end{equation}
This 2-D solution represents geometrically the $X-Y$ Heisenberg ferromagnet.This metric can easily be shown tobelong to the class of conical 2-D metrics since it is not globally flat.This can be accomplished by making use of the following coordinate transformation ${\rho}^{*}=e^{\frac{({\rho}+c)}{2}}$.Substitution of this transformation into the metric (\ref{25}) yields
\begin{equation}
ds^{2}=d{{\rho}^{*}}^{2}+{ln{\rho}^{*}}^{2}d{\beta}^{2}
\label{26}
\end{equation}
By expansion of the term ${ln{\rho}^{*}}$ around the radius ${{\rho}^{*}}_{0}$ we obtain the metric
\begin{equation}
ds^{2}=d{{\rho}^{*}}^{2}+\frac{{{\rho}^{*}}^{2}}{{{{\rho}^{*}}_{0}}^{2}}d{\beta}^{2}
\label{27}
\end{equation}
But this last metric can be transformed locally to a flat metric
\begin{equation}
ds^{2}=d{{\rho}^{*}}^{2}+{{\rho}^{*}}^{2}d{{\beta}^{*}}^{2}
\label{28}
\end{equation}
where the performed transformation ${{\beta}^{*}}=\frac{{\beta}}{{{\rho}^{*}}_{0}}$ produces a deficit angle carachteristic of the conic metrics since when the angle ${\beta}$ goes from $0$ to ${2{\pi}}$ the angle ${{\beta}^{*}}$ goes from zero to $\frac{2{\pi}}{{\rho}}$.Thus as we wish to show the Heisenberg ferromagnetic metric is not globally flat but in fact describes a conical defect.
\section{Geodesics around the ferromagnet}
From the above metric we obtain the following geodesics of a magnetic test particle which must be a magnetic monopole 
\begin{equation}
\frac{d^{2}x^{i}}{dt^{2}}+{\Gamma}^{i}_{jk}{\frac{dx^{j}}{dt}}{\frac{dx^{k}}{dt}}=0
\label{29}
\end{equation}
where ${\Gamma}^{i}_{jk}$ is the 2-D Christoffel-Levi-Civita connection of Riemannian geometry.The only nonvanishing Christoffel symbols are  
\begin{equation}
{\Gamma}^{1}_{11}=\frac{1}{2}
\label{30}
\end{equation}
and
\begin{equation} 
{\Gamma}^{2}_{21}=\frac{1}{{\rho}}
\label{31}
\end{equation}
and 
\begin{equation}
{\Gamma}^{1}_{22}=-{\rho}e^{-({\rho}+c)}
\label{32}
\end{equation}
Substitution of these equations into the geodesics equations yields
\begin{equation}
\ddot{{\rho}}+\frac{1}{2}{\dot{{\rho}}}^{2}-{\rho}e^{-({\rho}+c)}{\dot{{\beta}}}^{2}=0
\label{33}
\end{equation}
and  
\begin{equation}
{\ddot{{\beta}}}+\frac{\dot{\rho}}{{\rho}}{\dot{\beta}}=0
\label{34}
\end{equation}
Solution of equation (\ref{31}) is
\begin{equation}
{\dot{\beta}}={{\rho}}^{-1}
\label{35}
\end{equation}
Substitution of this equation into the geodesic equation (\ref{33}) one obtains
\begin{equation}
\ddot{{\rho}}+\frac{1}{2}{\dot{{\rho}}}^{2}-e^{-({\rho}+c)}\frac{1}{{\rho}}=0
\label{36}
\end{equation}
In the case ${\rho}$ approaches the ferromagnet the geodesic equation reduces to  
\begin{equation}
\ddot{{\rho}}+\frac{1}{2}{\dot{{\rho}}}^{2}-\frac{(1-{\rho})}{{\rho}}=0
\label{37}
\end{equation}
or
\begin{equation}
\ddot{{\rho}}+\frac{1}{{\rho}}=0
\label{38}
\end{equation}
Substitution of (\ref{38}) into {\ref{35}) one obtains the following solution
\begin{equation}
{\rho}={\beta}t+d
\label{39}
\end{equation}
which is a straight line and $d$ is an integration constant.Equation (\ref{38}) also tell us that the force between the Heisenberg ferromagnet and the magnetic monopole is attractive.
\section{Conclusions}
The metric for the Heisenberg ferromagnet in two dimensions was shown to be a Riemannian conical metric defect.Geodesics are given by straight lines in close analogy with the geodesics of the test particles around the dislocations in crystals investigated by F.Moraes \cite{8}and more recently by myself \cite{9} in the case of planar 3-D defects.Recently Azevedo,Furtado and Moraes \cite{10} have used the same conical geometry we investigated here to study repulsion of charges by positive curvature disclinations in monolayer graphite.It seems that together with our result here crystal geometry with defects are given by Riemann geometry in vacuum and by Riemann-Cartan geometry inside matter in analogy with the roles played by General Relativity and Einstein-Cartan theory in Gravitation. 
\section*{Acknowledgments}
I am very much indebt to                                                   S.Amari,K.Kondo,A.Wang,P.S.Letelier,C.Koehler, W.Kopczynski and P. von Ranke for helpful discussions on the subject of this paper.Financial support from Universidade do Estado do Rio de Janeiro (UERJ) and CNPq. is grateful acknowledged.


\newpage
\begin{thebibliography}{10}
\bibitem{1}H.Kleinert,Gauge Fields in Condensed Matter Physics,(1989),World scientific.
\bibitem{2}L.Pismen,Nonlinear Vortices:From Liquid Crystals to Superfluids.From Non-Equilibrium Patterns to Cosmic strings,(1999),Oxford Science Publications. 
\bibitem{3}L.C.Garcia de Andrade,Class. and Quantum Grav.(1999),16,6,2097.
\bibitem{4}P.S.Letelier and A.Wang,J.Math.Phys.(1995),36:3023. 
\bibitem{5}S.Amari,RAAG Mem.3(1962),257.
\bibitem{6}G.Asanov;Finsler Geometry,Relativity and Gauge Theories,(1985),Reidel.
\bibitem{7}L.C.Garcia de Andrade,Bloch walls with spin density as planar torsion defects,gr-qc/                 Los Alamos e-print archives
\bibitem{8}F.Moraes,Phys.Lett.A,(1995).
\bibitem{9}L.C.Garcia de Andrade,Phys.Lett.A,(1999)256.
\bibitem{10}S.Azevedo,C.Furtado and F.Moraes,Charge localization around disclinations in monolayer graphite,cond-mat/9606182.
\end{thebibliography}
\end{document}